\documentclass{article}

\usepackage{arxiv}

\usepackage[utf8]{inputenc} %
\usepackage[T1]{fontenc}    %
\usepackage{hyperref}       %
\usepackage{url}            %
\usepackage{booktabs}       %
\usepackage{amsfonts}       %
\usepackage{nicefrac}       %
\usepackage{microtype}      %
\usepackage{lipsum}
\usepackage{soul}
\usepackage{color}
\usepackage{float}
\usepackage{graphicx}
\usepackage{caption}
\usepackage{subcaption}
\usepackage{titlesec}
\usepackage{multirow}

\usepackage{amsmath}
\usepackage{soul,color}
\usepackage[section]{placeins}
\usepackage{array, makecell} %
\usepackage{float}
\usepackage[section]{placeins}
\usepackage{makecell}

\usepackage[nodisplayskipstretch]{setspace}

\title{AlphaCrystal-II: Distance matrix based crystal structure prediction using deep learning}

\author{
  Yuqi Song \\
  Department of Computer Science\\
  University of Southern Maine\\
  Portland, ME, 04131, USA \\
   \And
Rongzhi Dong, Lai Wei\\
  Department of Computer Science and Engineering\\
  University of South Carolina\\
  Columbia, SC, 29201, USA \\
    \And
 Qin Li\\
 College of Big Data and Statistics\\
  Guizhou University of Finance and Economics\\
  Guiyang China 550050 \\
   \And
 Jianjun Hu \thanks{Corresponding author: J.H. (http://www.cse.sc.edu/~jianjunh)}\\
  Department of Computer Science and Engineering\\
  University of South Carolina\\
  Columbia, SC, 29201, USA \\
  \texttt{jianjunh@cse.sc.edu} \\
}

\begin{document}
\maketitle

\begin{abstract}
Computational prediction of stable crystal structures has a profound impact on the large-scale discovery of novel functional materials. However, predicting the crystal structure solely from a material's composition or formula is a promising yet challenging task, as traditional ab initio crystal structure prediction (CSP) methods rely on time-consuming global searches and first-principles free energy calculations. Inspired by the recent success of deep learning approaches in protein structure prediction, which utilize pairwise amino acid interactions to describe 3D structures, we present AlphaCrystal-II, a novel knowledge-based solution that exploits the abundant inter-atomic interaction patterns found in existing known crystal structures.
AlphaCrystal-II predicts the atomic distance matrix of a target crystal material and employs this matrix to reconstruct its 3D crystal structure. By leveraging the wealth of inter-atomic relationships of known crystal structures, our approach demonstrates remarkable effectiveness and reliability in structure prediction through comprehensive experiments. This work highlights the potential of data-driven methods in accelerating the discovery and design of new materials with tailored properties.
\end{abstract}

\keywords{crystal structure prediction \and distance matrix \and deep learning \and residual neural network \and global optimization}

\section{Introduction}

Computational discovery of novel functional materials has enormous potential in transforming a variety of industries such as mobile communication, electric vehicles, quantum computing hardware, and catalysts\cite{oganov2019structure}. Compared to traditional Edisonian or trial-and-error approaches which usually strongly depend on the expertise of the scientists, computational materials discovery has the advantage of efficient search in the vast chemical design space. Among these methods, inverse design\cite{zunger2018inverse,kim2020inverse}, generative machine learning models\cite{dan2019generative,bradshaw2019model,kim2020inverse,noh2019inverse,ren2020inverse}, and crystal structure predictions \cite{glass2006uspex,oganov2011modern,kvashnin2019computational} are among the most promising approaches for new materials discovery.

Crystal structure prediction (CSP) is a notoriously hard problem \cite{maddox1988crystals,woodley2008crystal,lyakhov2013new} since a CSP algorithm has to find a crystal structure with the lowest free energy for given a chemical composition (or a chemical system such as Mg-Mn-O with variable composition) at given pressure-temperature conditions. CSP is highly desirable as the predicted crystal structure of a chemical substance allows many physicochemical properties to be predicted reliably by machine learning models or calculated routinely using first-principle calculation \cite{xie2018crystal,omee2022scalable}. It is assumed that lower free energy corresponds to more stable arrangements of atoms. The CSP approach for new materials discovery is especially appealing due to the efficient sampling algorithm that generates diverse chemically valid candidate compositions with low free energies\cite{dan2019generative,wei2022crystal}.  CSP algorithms based on evolutionary algorithms \cite{oganov2006crystal} and particle swarm optimization \cite{wang2015materials} have led to a series of new materials discoveries \cite{oganov2011evolutionary,oganov2019structure,wang2020calypso}. However, these global free energy search-based algorithms have a major obstacle that limits their successes to relatively simple crystals \cite{oganov2019structure,zhang2017materials} (mostly binary materials with less than 20 atoms in the unit cell\cite{oganov2019structure,wang2020calypso}) due to their dependence on the costly density functional theory (DFT) calculations of free energies for sampled structures. With a limited DFT calculations budget, how to efficiently sample the atom configurations becomes a key issue \cite{oganov2011evolutionary,lyakhov2013new}. To improve the sampling efficiency, a variety of strategies have been proposed such as exploiting symmetry\cite{pretti2020symmetry} and pseudosymmetry\cite{lyakhov2013new}, smart variation operators, clustering, and machine-learning interatomic potentials with active learning \cite{podryabinkin2019accelerating}. However, the scalability of these approaches remains an unsolved issue.

With the mature development of deep learning techniques, several studies have applied those novel methods in the materials science field \cite{agrawal2019deep}, particularly in the areas of material property prediction \cite{louis2020graph}, material discovery \cite{song2021computational}, and material design\cite{kim2021deep}. %
Recently, several emerging works have utilized deep learning methods to predict material structures. Ryan et al. \cite{ryan2018crystal} reconstructed the crystal structure prediction problem as predicting the likelihoods of particular atomic sites in the structure. Their trained model successfully distinguishes chemical components based on the topology of their crystallographic environment. They use the model to analyze templates derived from the known crystal structures in order to predict the likelihood of forming new compounds. Cheng et al. \cite{cheng2022crystal} recently proposed the GNOA framework that employs a graph network model to connect crystal structures and their formation enthalpies and then merged this model with an optimization algorithm for CSP. Although they successfully predicted 29 crystal structures, the main limitation of the study is the failure to predict structures of complicated chemical formulas. Machine learning based and heuristic rules based template methods have also been proposed for CSP prediction with strong performance \cite{CSPML,wei2022tcsp}.

Our work here is inspired by the great success of AlphaFold \cite{jumper2021highly} in protein structure prediction since crystal structure prediction in material and protein structure prediction (PSP) in bio-informatics share many similarities \cite{agrawal2019deep}. (1) PSP aims to construct the three-dimensional (3D) protein structure from a protein only given its amino acid sequence \cite{wei2019protein}, while CSP seeks to identify a crystal structure with the lowest free energy for certain chemical compositions. (2) Features and functions of a given protein and material can be explored more effectively if we uncover their structures. (3) Bboth tasks are typically time-consuming and costly when relying solely on laboratory experiments or first-principles calculations.
Motivated by the recent breakthrough of deep learning in PSP with contact map \cite{adhikari2018dncon2} and DeepMind's AlphaFold \cite{jumper2021highly}, we investigate the recent research on deep learning based protein structure prediction \cite{emerson2017protein,kuhlman2019advances,ingraham2019generative}, in which the contact map places a key role in their algorithm design.  
For a protein 3D structure, a contact map is a binary form of the distance matrix with a distance threshold, where the distance is calculated between every pair of residues  \cite{emerson2017protein}. Adhikari et al.\cite{adhikari2018dncon2} proposed DNCON2, an \emph{ab initio} protein contact map predictor based on two-level deep convolutional neural networks: at the first level, they used five convolution neural networks (CNNs) with multiple-distance thresholds to predict preliminary contact probabilities as the additional features in the next step; then, the another CNN was used to predict the final contact probability map. Since a distance map can reveal more structural information than a contact map, in \cite{senior2020improved}, they presented the AlphaFold, a deep learning model that achieved high accuracy even when there were fewer homologous sequences available. 
Taking inspiration from AlphaFold's success in protein structure prediction and acknowledging the similarities between crystal structure prediction and protein structure prediction as previously discussed, we aim to exploit the relationship between atomic pairs for crystal structure prediction. Unlike protein structure prediction, where co-evolution or correlated mutation relationships among amino acids are leveraged, the concept of such relationships among atoms is not well-established in the field of crystal structure prediction (CSP). Moreover, there is no existing method for predicting the distance matrix of crystal structures from compositions only except our previous work \cite{hu2023deep}, in which the binary contact map is predicted and exploited for CSP with moderate success. 

A large number of physical and chemical rules govern the interactions and bonding between atoms, leading to the predictability of the distance matrix for crystal structures from the composition. Building upon this foundation, we propose a distance matrix-based model and algorithm for CSP, the block diagram of which is illustrated in Figure \ref{fig:framework}. Our work demonstrates the possibility of making accurate predictions of the crystal structure for a given composition based on the distance matrix predicted by a deep neural network.
Drawing inspiration from the successful strategies employed in protein structure prediction and leveraging the inherent physical and chemical principles governing atomic interactions, our approach provides a novel perspective on crystal structure prediction. The distance matrix, which encapsulates the intricate relationships between atoms, serves as a powerful representation for predicting the three-dimensional arrangement of atoms in a crystal structure. This work opens up new avenues for exploring data-driven and knowledge-based methods in the field of materials science, with the potential to accelerate the discovery and design of new materials with tailored properties.

Our contributions can be summarized as follows:

\begin{itemize}
  \item We proposed a data-driven AlphaCrystal-II model for CSP which exploits the abundant atomic pair (bond) knowledge and utilizes the trained deep residual neural networks for distance matrix prediction and crystal structure reconstruction.
  
  \item We conducted extensive experiments to demonstrate AlphaCrystal-II's competitive effectiveness and reliability in crystal structure prediction.
  
  \item We compared AlphaCrystal-II with the GNOA algorithm, a modern machine learning potential based CSP method. We found that AlphaCrystal-II can achieve better performance in most cases and works not only for materials with simple compositions but also those materials with complex compositions.

  \item We evaluated the performance of the distance matrix prediction and structure reconstruction by the distance matrix prediction network and the genetic algorithm DMCrystal, which showed a strong performance that explains the good performance of AlphaCrystal-II.
\end{itemize}

\section{Methods}
\subsection{Framework of AlphaCrystal-II model for crystal structure prediction}

Given a material composition, a simple yet effective approach for crystal structure prediction is to use the composition to find a possible template structure and conduct CSP using elemental substitution, which has led to powerful template based CSP algorithms such as CSPML\cite{CSPML} and TCSP\cite{wei2022tcsp}. The basic assumption here is that materials with similar compositions should have similar structures, which also implies that composition should contain enough information to predict the atomic pairwise contacts/bonds or distance matrices. This can be achieved by exploiting the composition similarity between the query sample and the training samples, the complementary bonding relationships of anions or cations, the oxidation state valence electron distributions, or the bonding preferences for a given pair of atoms. Although there is no co-evolutionary information or ancestral relationship among the protein structures, unlike in protein structures, the distance matrix based CSP is also feasible and promising.

\begin{figure}[ht]
  \centering
  \includegraphics[width=\linewidth]{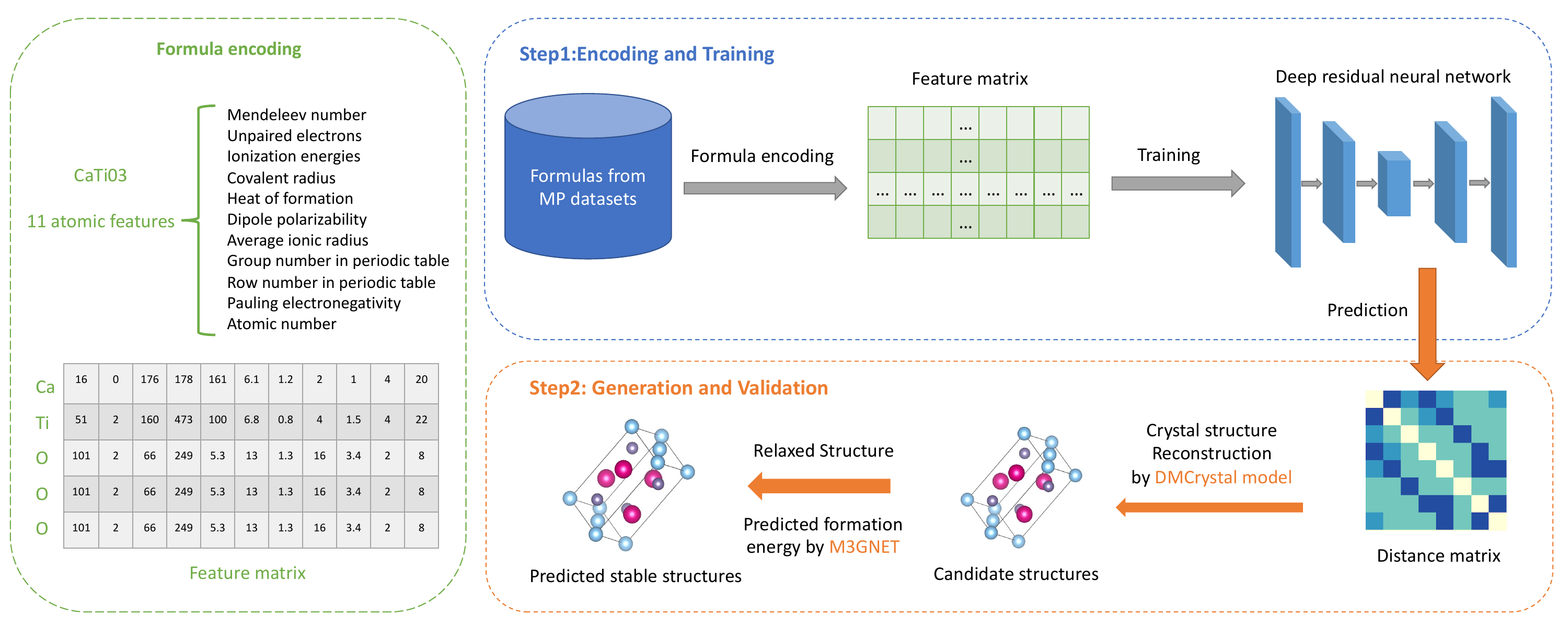}
  \caption{The AlphaCrystal-II framework for distance matrix-based crystal structure prediction. First, material compositions are encoded into feature matrices according to their atom features, which are used to train the deep neural network for predicting distance matrices. Next, the crystal structures are reconstructed by DMCCrystal GA along with predicted unit cell information. The resulting candidate structures are then relaxed by the M3GNet model. For multiple candidate structures, the top K structures are selected based on their predicted formation energies determined by M3GNet.}
  \label{fig:framework}
\end{figure}

Based on the above ideas and the great success of AlphaFold, we propose the AlphaCrystal-II model crystal structure prediction tasks. The framework of our model is shown in Figure \ref{fig:framework}. It contains an encoding module, a distance matrix prediction neural network, and a structure generation/reconstruction module. For a given material composition or formula, a feature matrix is constructed using the 11 elemental properties as shown in Table \ref{tab:11features}. This feature matrix of the composition is then fed to the deep residual network for predicting its distance matrix. The detailed architecture of this network model is shown in Figure \ref{fig:network}. The model is trained using known crystal structures selected from the Materials Project database \cite{jain2013commentary}. We also need to use the third-party machine learning model MLatticeABC to predict the lattice parameters and space groups \cite{li2021mlatticeabc}. The predicted distance matrix, along with the unit cell information, is used by the DMCrystal \cite{hu2020distance} genetic algorithm to reconstruct the final structure. Finally, the Bayesian Optimizer of the M3GNet \cite{chen2022universal} package is used to relax the predicted structures and estimate their formation energies using the M3GNet ML potential, thereby enabling the identification of stable structures.

In the sections below, we describe the details of the components of our AlphaCrystal-II modules.

\subsection{Feature matrix encoding}

Representing material composition information in an appropriate format to learn their atomic interactions is an essential and crucial preparation step for the AlphaCrystal-II algorithm because atomic interactions determine its final structure. Previous research has demonstrated that many atomic characteristics are related to ionic bonds, such as ionic bonding are electronegativity difference, ionization energy, electron affinity, atom size, ionic charge, lattice energy, polarizability, coordination number, and environmental conditions like temperature, pressure, and presence of other substances \cite{housecroft2012inorganic}. In our approach, we use 11 atomic features to encode the physicochemical information of each atom in the material, serving as input for the deep neural network model utilized for distance matrix prediction.

Specifically, each element/atom is represented by 11 chemical descriptors (see Table \ref{tab:11features} for details) including Mendeleev Number \cite{villars2004data}, unpaired electrons, ionization energies, covalent radius, heat of formation, dipole polarizability \cite{schwerdtfeger20192018}, average ionic radius, group number and row number in the periodic table, pauling electronegativity, and atomic number. These 11 atomic features are represented by each row of the feature matrix as shown in Figure \ref{fig:framework}.

Each row of the feature matrix represents an element symbol in the unit cell, where $L$ is the number of atoms. We set the maximum number of atoms to be 12 in this study for simplicity, and if the number of atoms is less than 12, the remaining empty positions will be padded with zeros. Hence, the dimension of each feature matrix is $12 \times 11$.

\begin{table}[ht]
\centering
\caption{11 elemental features for atomic composition encoding}
\label{tab:11features}
\begin{tabular}{|l|l|}
\hline
Feature                        & Description \\ \hline
Mendeleev number (MN) \cite{villars2004data}           &   \makecell*[l]{Listing of chemical elements by column through the periodic system,  which is \\ effectively used to classify chemical systems.}         \\ \hline

Unpaired electrons             &  Electrons that occupy an orbital of an atom singly.             \\ \hline
Ionization energies            & The amount of energy required to remove an electron from an isolated atom or molecule.       \\ \hline
Covalent radius                &  Half of the distance between two atoms bonded covalently.           \\ \hline
Heat of formation              & \makecell*[l]{When one mole of a compound is produced from its basic elements, each substance is \\ in its normal physical state, the quantity of heat received or released.}        \\ \hline
Dipole polarizability \cite{schwerdtfeger20192018}  & \makecell*[l]{Describes the linear response of an electronic charge distribution with respect to an \\ externally applied electric field.}          \\ \hline
Average ionic radius       &    Average of a monatomic ion's radius in an ionic crystal structure.         \\ \hline
Group number in periodic table & \makecell*[l]{The number of valence electrons of the elements in a certain group, a group is a vertical \\ column of the periodic table.}       \\ \hline
Row number in periodic table   &  The number of rows of the element in the periodic table.       \\ \hline
Pauling electronegativity      & The power of an atom in a molecule to attract electrons to itself.         \\ \hline
Atomic number                  & The charge number of an atomic nucleus.         \\ \hline
\end{tabular}

\end{table}

\subsection{Deep residual network model for distance matrix prediction}

For distance prediction, we train a deep residual neural network to capture and exploit the intricate bonding interactions between atoms. This method leverages the vast distribution of inter-atomic interactions present in the large number of known crystal structures used as our training set, thereby enabling highly accurate pairwise distance predictions.

\begin{figure}[ht]
  \centering
  \includegraphics[width=0.85\linewidth]{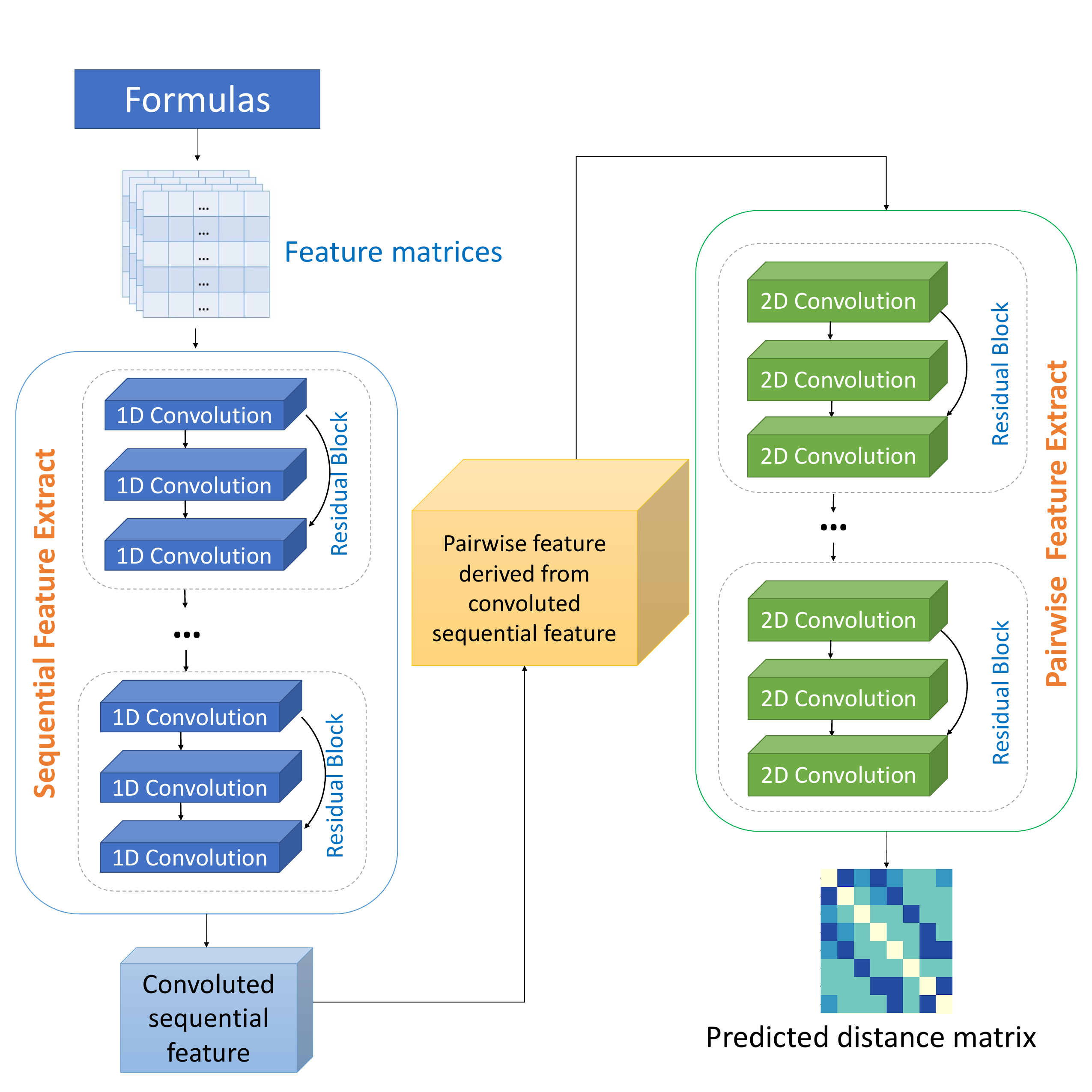}
   \caption{Deep neural network model for distance matrix prediction.}
  \label{fig:network}
\end{figure}

Figure \ref{fig:network} illustrates our deep neural network model, which comprises three main components. The first part consists of a sequence of stacked 1-dimensional (1D) residual network layers designed to learn the intricate features of atomic sites. In the second part, a 2-dimensional (2D) pairwise feature matrix is derived from the output of the 1D convolutional network through an outer product operation. Subsequently, we merge the convoluted sequential 
features and pairwise features to serve as the input for the next module. The third part encompasses a series of 2D residual network layers, which predict the distances between pairs of atoms, ultimately generating the predicted distance matrix.

In our studies, we set the maximum number of atoms in a formula to $L$ to accommodate the variable sizes of different crystal structures. In the experiments, $L$ is set to 12 (the same as the feature matrix dimension). For formulae with fewer atoms, we create the corresponding tensors by padding them with zeros.

The Residual Neural Network (ResNet) \cite{he2016identity} is a type of neural network architecture that stacks residual blocks on top of each other, forming a deep network through skip connections. This design has profoundly influenced the way deep neural networks are constructed. The skip connections between layers add the outputs from previous layers to the outputs of stacked layers, enabling the training of much deeper networks than what was previously possible.
In the AlphaCrystal-II model, we design two residual network modules: one module aims to extract sequential features, while the other module intends to derive pairwise features. Each residual network block comprises two convolutional layers, a batch normalization layer, and two nonlinear transformations. Our main architecture employs 9 building blocks for each module. The number of filters is doubled every 3 blocks, with the initial number of filters set to 32 and 256 for the first and second modules, respectively.
The residual network architecture, with its skip connections and residual blocks, enables the training of deeper neural networks by mitigating the vanishing gradient problem. This design choice allows our model to effectively capture the intricate patterns and interactions present in the atomic structures, leading to more accurate predictions of pairwise distances and ultimately improving the overall performance of our crystal structure prediction approach.

Since the discretized distance map is a binary matrix, we use the cross-entropy loss as the loss function for neural network training. It is defined as: 

\begin{equation} L_{crossEntropy} = - \frac{1}{N}  \sum_{i=0}^{N} y_i \cdot log (\hat{y}_i)  + (1 -y_i) \cdot log (1 -  \hat{y}_i) 
\end{equation}

Where $N$ is the maximum length of the formula, which is set to $12 \times 5$ and $12 \times 10$  in our experiments; $y_i$ is the true distance matrix label at position $i$, and $ \hat{y}_i$ is the predicted label at position $i$.

\subsection{3D crystal structure reconstruction}

Once the neural network predicts the distance matrix for a given composition, the crystal structure can be reconstructed using a genetic algorithm such as DMCCrystal \cite{hu2020distance}. This work has demonstrated that, given the pairwise atomic distance matrix with the space group, lattice parameters, and stoichiometry, the genetic algorithms can reconstruct the crystal structure that is close to the target crystal structure. For further improvement, these predicted structures can be used to seed the costly free-energy minimization-based ab initio CSP algorithms, as well as to obtain a more accurate crystal structure of certain components by DFT-based structural relaxation.

In this work, the reconstructed structures by the genetic algorithm are all relaxed using the Bayesian optimization algorithm with the M3GNet model \cite{chen2022universal}, which is a universal inter-atomic potential for materials based on graph neural networks with three-body interactions. It is trained by a lot of structural relaxations performed from the Materials Project for 89 elements of the periodic table with low energy, force, and stress errors. It has a wide range of uses in structural relaxation and dynamic simulations of materials across diverse chemical spaces.

\subsection{Evaluation metrics}
In assessing the performance of our distance map based structure prediction algorithm for crystal structure prediction (CSP), we use eight different metrics, including superpose distance, energy distance, Sinkhorn distance, fingerprint distance, OFM distance  \cite{wei2024towards}, Wyckoff MSE, Wyckoff RMSE, and distance map overlap.

The Wyckoff MSE and RMSE measure the similarity between the predicted structure and the true target structure. Distance overlap evaluates the precision of the predicted distance map compared to the target distance map. The equations for such metrics are shown in the following equations:

\begin{equation}
        \begin{aligned}
\mathrm{MSE} &=\frac{1}{n} \sum_{i=1}^{n}(y_{i}-\hat{y}_{i})^2 \\
\end{aligned}
\end{equation}

\begin{equation}
    \begin{aligned}
\mathrm{RMSE} &=\sqrt{\frac{1}{n} \sum_{i=1}^{n}(y_{i}-\hat{y}_{i})^{2}} 
\end{aligned}
\end{equation}

where $n$ is the number of independent atoms in the target crystal structure, $y_i$ and $\hat{y}_i$ are the coordinates of the corresponding atoms in the predicted and the target crystal structures. 

\begin{equation}
\operatorname{DM overlap}=\frac{|TargetDM_{atom~exist} \cap PredictedDM_{atom~exist}|}{|Target~atoms|}
\end{equation}

where $TargetDM$ is the target distance map and $PredictedDM$ is the predicted distance matrix of a given composition, both only contain 1/0 entries. $Target~atom$ means the total number of atoms in the target distance map.
$TargetDM~atoms_{exist} \cap PredictedDM~atom_{exist}$ denotes the common existent atoms of $TargetDM$ and $PredictedDM$.
This performance metric evaluates the overlap between two distance map matrices, with values ranging from 0 to 1, where 1 indicates perfect overlap. This metric, also known as distance map accuracy, quantifies the fidelity of the predicted distance map. To obtain the predicted distance map, we use a threshold of 8 amstrong as the cutoff to convert a given distance matrix into a binary distance map.

\section{Results}
\label{sec:headings}

\subsection{Dataset}
We collected material ids, formulas, and crystal structures (cif files) from the Materials Project database \cite{jain2013commentary} by Pymatgen API \cite{ong2013python}. Filtering out compounds with a maximum of 12 atoms in the formula yielded 18,800 compounds from the Materials Project, constituting our initial dataset named Mp\_12. Furthermore, to compare our prediction results with the GNOA algorithm \cite{cheng2022crystal}, we removed 29 formulas that predicted structures with high performance from the training set, therefore, there are 18,776 compounds in Mp\_12 datasets. Moreover, we isolated materials with a cubic crystal system to create the Mp\_12\_cubic dataset. Additionally, we extracted formulas comprising binary and ternary elements to form the Mp\_12\_binary and Mp\_12\_ternary datasets, respectively. Subsequently, we trained four distance matrix prediction models using each of these four datasets and evaluated their performance.

\begin{table}[!htb]
\begin{center}
\caption{Datasets}
\label{tab:dataset}
\begin{tabular}{lll}
\hline
Dataset             & Number of samples & Description \\ \hline
Mp\_12    & 18,776  &  \makecell*[l]{materials with $\leq 10$ atoms per unit cell picked from Materials Project \\and after the 29 test samples in the GNOA study are removed}    \\
Mp\_12\_cubic & 3,298           &   cubic materials extracted from Mp\_12 \\
Mp\_12\_binary & 5,848           &   binary materials extracted from Mp\_12      \\
Mp\_12\_ternary & 11,615           &   ternary materials extracted from Mp\_12     \\
\hline
\end{tabular}
\end{center}
\end{table}

\begin{figure}[ht]
  \centering
  \includegraphics[width=0.85\linewidth]{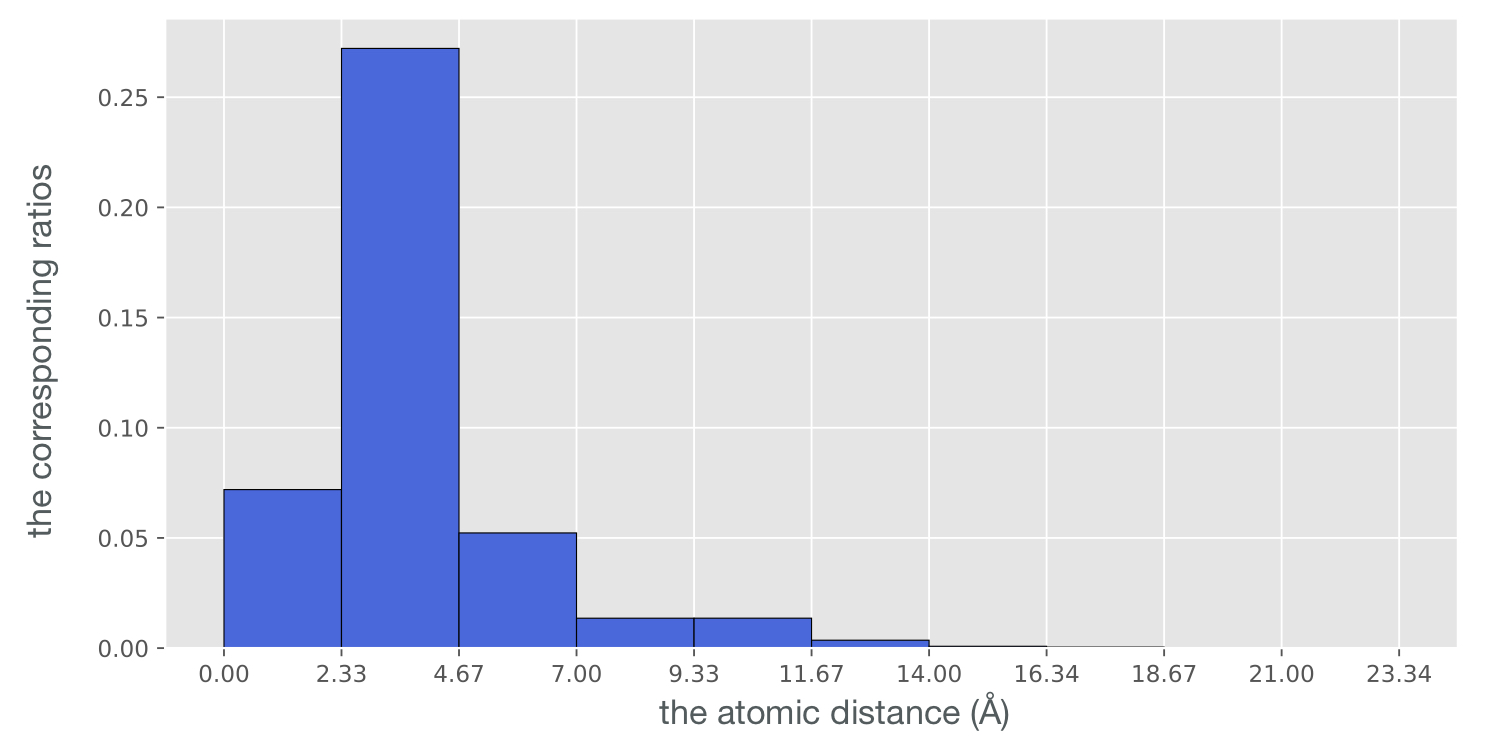}
  
  \caption{Overall atomic distance distribution of Mp\_12 dataset. The x-axis represents the atomic distance and the y-axis represents the corresponding ratios. The most atomic distances are in the range of 2.33-4.67, accounting for more than 25\% and there are just a few distances more than 14.}
  \label{fig:distance_distribution}
\end{figure}

\paragraph{Distance discretization} 
We counted the atomic distances of 18,776 samples in the Mp\_12 dataset, the overall distribution is shown in Figure \ref{fig:distance_distribution} where the smallest distance is 0.9488, the largest distance is 23.3361, and the majority of distances are between 0 to 7 groups. To facilitate further analysis, we divided the continuous distance values with equal width and store them in several groups by equation \ref{con:width}. 

\begin{equation}
Interval~width = (Maximum\_value - Minimum\_value) / N
\label{con:width}
\end{equation}

where $N$ represents the number of groups, we set it to 5, 10, or 20 in experiments. Furthermore, we use the one-hot encoding to get the expression of each value. This method turns the regression problem into a classification problem by turning continuous distances into discrete values. Therefore, the cross-entropy loss can be used as the loss function for training neural networks, which makes the prediction more accurate.

For each material in our dataset, we encoded 11 atomic features to a 12*11 matrix (the maximum atomic number in the unit cell is 12). After one-hot encoding, the dimension of the distance matrix is 12*(12*N). In our experiments, the training samples are randomly selected with different sample sizes according to the needed dataset sizes, such as 1,000 or 10,000 samples; after model training, 100 or 1000 samples that are not included in the training sets are selected for CSP algorithm testing and performance analysis. 

\subsection{Prediction performance of AlphaCrystal-II}

\subsubsection{CSP Performance Comparison of AlphaCrystal-II model and GNOA model}

We first evaluated the CSP performance of our AlphaCrystal-II and compared it with the other machine learning potential based CSP algorithm, GNOA \cite{cheng2022crystal}. 
GNOA uses a graph network energy potential (MEGNet) \cite{chen2019graph} and an optimization algorithm such as random search, Bayesian Optimization, and Particle Swarm Optimization for crystal structure prediction. The graph neural network is trained to build a correlation model between crystal structures and their formation enthalpies, which is utilized by an optimization algorithm to search for the crystal structure that has the lowest formation enthalpy. Their report \cite{cheng2022crystal} showed 29 selected compounds with good prediction performance. However, their selected test samples are almost all belonging to simple binary compounds with few atoms within the unit cell. To more objectively compare the predicted performance of GNOA and AlphaCrystal-II, we evaluated these two models by predicting 30 diverse material structures separately, and the experimental results are summarized in Table \ref{tab:perf_compare}.

We randomly selected 10 binary, ternary, and quaternary materials from our MP\_12 dataset, respectively. For each formula, we used both GNOA and AlphaCrystal-II to predict the probable crystal structures and then calculated the superpose distance between each predicted structure and the ground truth structure.
Our AlphaCrystal-II can generate many structure candidates all at once. However, GNOA can only provide one structure that has the lowest formation energy. Thus, we compare GNOA's best structure with AlphaCrystal-II's Top-10 and Top-20 structure candidates.

As shown in Table \ref{tab:perf_compare}, for all 30 formulas, AlphaCrystal-II's Top-10 structure candidates achieved the best performance 17 times, and Top-20 candidates expanded this number to 23. In contrast, GNOA only outperformed AlphaCrystal-II in 7 formulas.
In two of these 7 formulas, BePd and PaIn$_3$, GNOA's performance only outperformed AlphaCrystal-II by 0.0004 and 0.0002, respectively. This means that AlphaCrystal-II's structure had comparable properties.
Furthermore, GNOA encountered challenges in handling complex formulas, particularly for quaternary materials. It failed to predict the structures of Rb$_2$LiRhCl$_6$, Y$_3$Al$_3$NiGe$_2$, and LiFe$_2$(ClO)$_2$. In contrast, AlphaCrystal-II was able to provide many structure candidates that were similar to the ground truth structures.
Moreover, for two formulas, Ce$_3$Pm and DyThCN, AlphaCrystal-II could generate structures that were the same as the ground truth structures, which demonstrates the power of our AlphaCrystal-II.

\begin{table}[h]
\begin{center}
\caption{Performance Comparison of AlphaCrystal-II and GNOA.}
\label{tab:perf_compare}
\begin{tabular}{|c|c|c|c|c|c|}
\hline
Type              & Formula & Mp\_id     &  \makecell {GNOA \\ superpose\\distance(Å) } & \makecell { AlphaCrystal-II\\ Top-10 \\superpose \\distance(Å) }& \makecell {AlphaCrystal-II\\ Top-20 \\superpose\\distance(Å)} \\ \hline

\multirow{9}{*}{Binary} & TiIn & mp-1216825  & 0.7775
 & \textbf{0.0158}  & \textbf{0.0158}

 \\ \cline{2-6} 
& ScAl   & mp-331       & 1.4594
  & \textbf{0.0032}
      & \textbf{0.0032}
       \\ \cline{2-6} 
& Tm$_3$P   & mp-971958     &  1.9484
  & \textbf{0.5625} & \textbf{0.5625}
 \\ \cline{2-6} 
& BePd   & mp-11274    &  \textbf{0.0085}
       & 0.0089
   & 0.0089
      \\ \cline{2-6} 
& PaIn$_3$  & mp-861987  &  \textbf{1.6442}
   & 1.6644
   & 1.6644
    \\ \cline{2-6} 
& Ce$_3$Pm  & mp-1183767    &  0.0456
       & \textbf{0}      & \textbf{0}       \\ \cline{2-6} 
& GdCo$_5$  & mp-1077071   &  0.8275
  & \textbf{0.024} & \textbf{0.024}
 \\ \cline{2-6} 
& LiC$_6$   & mp-1001581  &  1.1425
  & \textbf{0.843}
 & \textbf{0.843}
 \\ \cline{2-6} 
& BaAu$_5$  & mp-30364    & \textbf{1.0055}
  & 1.265
 & 1.265
  \\ \cline{2-6} 
& Zr$_4$Al$_3$ & mp-12752   &  1.5999
  & \textbf{0.0204}   & \textbf{0.0204}
 \\ \hline

\multirow{9}{*}{Ternary} & TmIn2Sn & mp-1216827  & 1.6361
 &  1.1517 & \textbf{0.9419}
 \\  \cline{2-6}  
& CeAlO$_2$   & mp-1226604    & 1.5539
    & 1.277
  & \textbf{1.1403}
 \\ \cline{2-6}  
& ThBeO$_3$    & mp-1187421   & \textbf{1.0304}
    & 1.2676
   & 1.2676
  \\ \cline{2-6}
& Zn(CuN)$_2$  & mvc-15351   & \textbf{0.6701}
    & 1.0663
   & 1.0663
  \\ \cline{2-6}
& LiVS$_2$     & mp-7543     & 1.7123
    & \textbf{0.7939}
   & \textbf{0.7939}
  \\ \cline{2-6}
& LiTiTe$_2$   & mp-10189     & 1.2147
    & \textbf{1.0687}
   & \textbf{1.0687}
   \\ \cline{2-6}
& Sr(AlGe)$_2$ & mp-1070483  & 1.3211
    & \textbf{0.2149}
 & \textbf{0.2149}
  \\ \cline{2-6}
& NdTiGe    & mp-22331     & 1.9989
    & 1.5848
 & \textbf{1.4723}
\\ \cline{2-6}
& Mn$_4$AsP$_3$   & mp-1221760   & 1.596
    &  \textbf{1.3444}
   & \textbf{1.3444}
 \\ \cline{2-6}
& Mn$_3$FeP$_4$   & mp-1221749    & 1.1733
    & 1.0396
   & \textbf{0.9941}
 \\ \hline

\multirow{9}{*}{Quaternary} & DyThCN  & mp-1225528  &  1.3307
 & 0.4564 & \textbf{0}

 \\ \cline{2-6}
& EuNbNO$_2$       & mp-1225127    & 1.0466
  & \textbf{0.621}
  & \textbf{0.621}
 \\  \cline{2-6}
& LaNiPO       & mp-1079685     & 2.0049
  & 1.4417
  & \textbf{1.3439}
 \\ \cline{2-6}
& SrFeMoO$_5$      & mp-690817       & \textbf{0.4081}  & 1.8519
  & 1.1096
 \\ \cline{2-6}
& FeCu$_2$SnS$_4$    & mp-628568    & 2.1553
  & 1.5029
  & \textbf{1.2196}
 \\ \cline{2-6}
& LiTb(CuP)$_2$    & mp-8220       & 1.5779
  & 0.9468
  & \textbf{0.4403}
  \\ \cline{2-6}
& Rb$_2$LiRhCl$_6$    & mp-1206187      & None    & \textbf{0.6224}
  & \textbf{0.6224}
  \\ \cline{2-6}
& Y$_3$Al$_3$NiGe$_2$    & mp-10209      & None    & \textbf{1.1824}
  & \textbf{1.1824}
  \\ \cline{2-6}
& LiFe$_2$(ClO)$_2$   & mp-755254      & None    & \textbf{0.9999}
   & \textbf{0.9999}
 \\ \cline{2-6}
& Mg$_2$VWO$_6$      & mp-1303315     & \textbf{1.5126}
  & 1.5777
  & 1.5706
  \\ \hline
& \# of best       &    & \textbf{7}  & \textbf{17}  & \textbf{23}  \\ \hline

\end{tabular}
\end{center}
\end{table}

\begin{figure}[htb!]
    \centering
    \includegraphics[width=\textwidth]{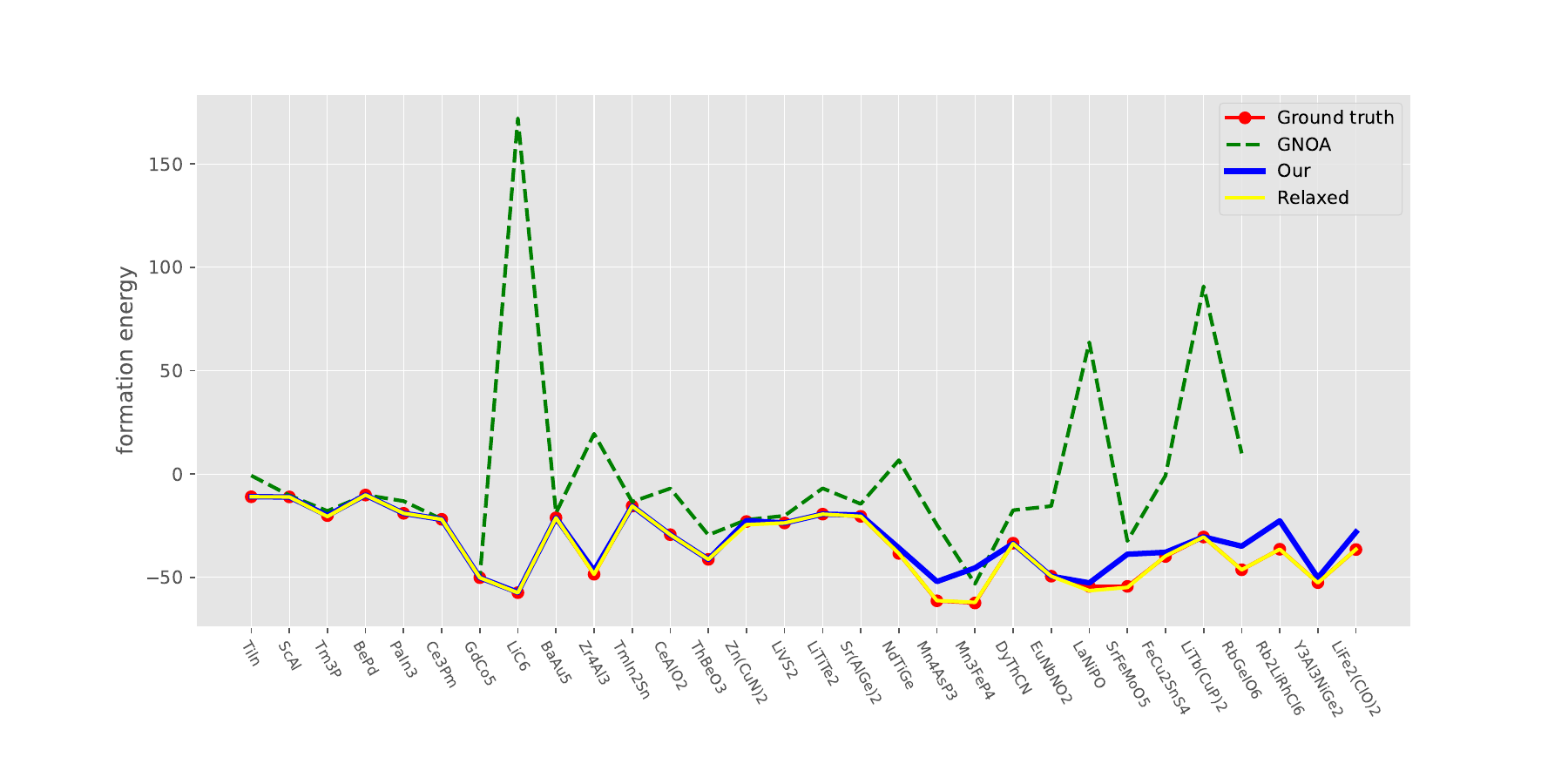}
    \caption{Comparison of formation energies of predicted structures by GNOA and AlphaCrystal-II.}
    \label{fig:compare_eform}
\end{figure}

In addition, formation energy serves as a crucial criterion for assessing the stability of crystal structures. Hence, we compare the lowest formation energies of existing crystal structures with those of our predicted structures generated by M3GNet, as depicted in Figure \ref{fig:compare_eform}. From this figure, we can observe that the formation energies of predicted structures by AlphaCrystal-II closely align with those of the target structures, in contrast to the results obtained by GNOA, which further proves the effectiveness of our method.

\begin{figure}[htb!]
	\centering
	\begin{subfigure}{.33\textwidth} %
		\includegraphics[width=\textwidth]{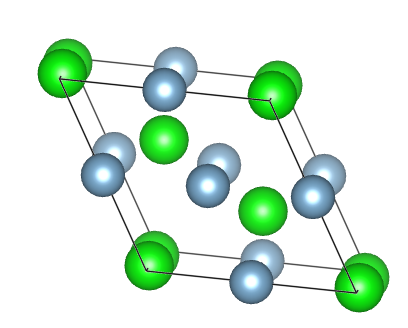}
		\caption{}
		\vspace{3pt}%
	\end{subfigure}
	\begin{subfigure}{.33\textwidth}
		\includegraphics[width=\textwidth]{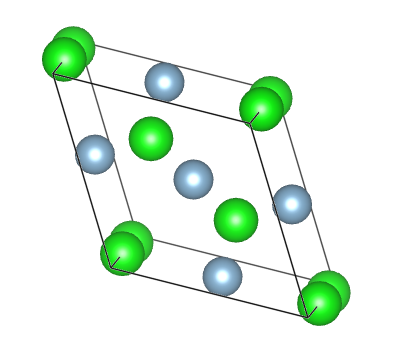}
		\caption{}
		\vspace{3pt}%
	\end{subfigure}
 	\begin{subfigure}{.33\textwidth}
		\includegraphics[width=\textwidth]{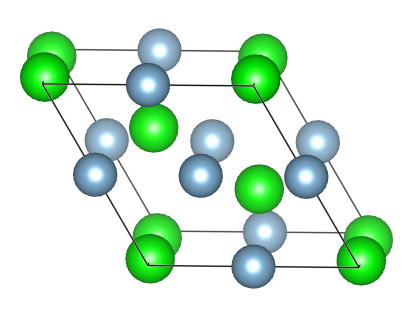}
		\caption{}
		\vspace{3pt}%
	\end{subfigure}

	\begin{subfigure}{.33\textwidth}
		\includegraphics[width=\textwidth]{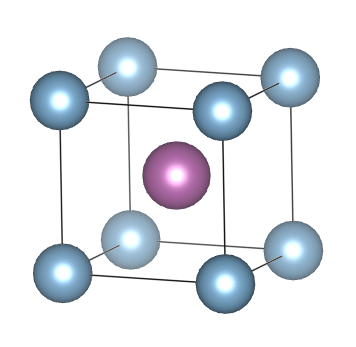}
		\caption{}
	\end{subfigure}
	\begin{subfigure}{.33\textwidth}
		\includegraphics[width=\textwidth]{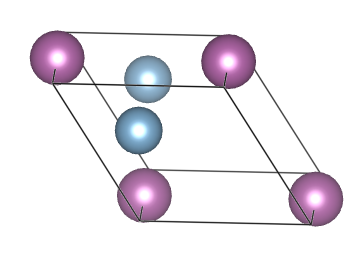}
		\caption{}
	\end{subfigure}
     \begin{subfigure}{.33\textwidth}
		\includegraphics[width=\textwidth]{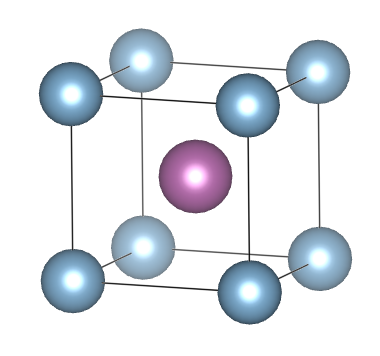}
		\caption{}
	\end{subfigure}

 	\begin{subfigure}{.33\textwidth}
		\includegraphics[width=\textwidth]{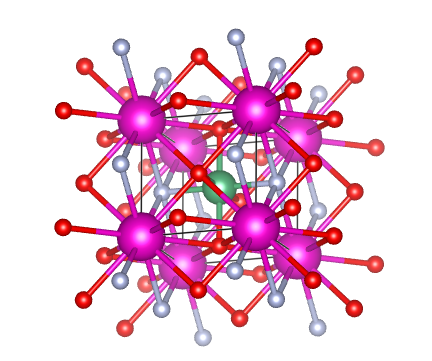}
		\caption{}
	\end{subfigure}
	\begin{subfigure}{.33\textwidth}
		\includegraphics[width=\textwidth]{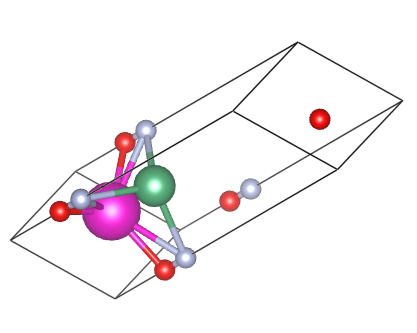}
		\caption{}
	\end{subfigure}
     \begin{subfigure}{.33\textwidth}
		\includegraphics[width=\textwidth]{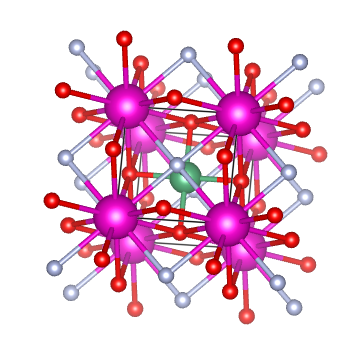}
		\caption{}
	\end{subfigure}

	\caption{Examples of predicted crystal structures by AlphaCrystal-II. (a) the ground truth structure of Zr$_{4}$Al$_{3}$; (b) the predicted structure of Zr$_{4}$Al$_{3}$ by GNOA; (c) the predicted structure of Zr$_{4}$Al$_{3}$ by AlphaCrystal-II; (d) the ground truth structure of ScAl; (e) the predicted structure of ScAl by GNOA; (f) the predicted structure of ScAl by AlphaCrystal-II; (g) the ground truth structure of EuNbNO$_{2}$; (h) the predicted structure of EuNbNO$_{2}$ by GNOA; (i) the predicted structure of EuNbNO$_{2}$ by AlphaCrystal-II; }
	\label{fig:predictedstructures}
\end{figure}

We further check three case studies for the prediction performance of these two algorithms. Three samples of predicted crystal structures by GNOA and AlphaCrystal-II and their ground truth (target) structures are shown in Figure \ref{fig:predictedstructures}. 
For compound Zr\textsubscript{4}Al\textsubscript{3}, the RMSE of the distance matrix is 0.4818 \AA and the RMSE of the predicted structure is 0.2183 \AA.

\subsubsection{Performance Comparison of AlphaCrystal-II against AlphaCrystal}

Here we evaluate how the distance matrix based AlphaCrystal-II performs compared to the atomic map based AlphaCrystal-I \cite{hu2023deep}. We used both algorithms to predict the crystal structures for the following test samples from the Materials Project database, including Pb$_{4}$O$_{4}$, Co$_{4}$P$_{8}$, Ir$_{4}$N$_{8}$, V$_{2}$Cl$_{10}$, Co$_{2}$As$_{2}$S$_{2}$, 
V$_{4}$O$_{4}$F$_{4}$, Fe$_{4}$As$_{4}$Se$_{4}$, Mn$_{4}$Cu$_{4}$P$_{4}$. We then compare their performance metrics using four metrics selected from the CSBenchMetrics \cite{wei2024towards}. The results are shown in Table \ref{tab:compareI_and_II}.

For eight selected formulas, we used both AlphaCrystal-I and AlphaCrystal-II to generate a branch of probable structures and then chose the top 20 with the lowest formation energy value as final structure candidates. Then, we employed five different metrics including energy distance, Sinkhorn distance, superpose distance, fingerprint distance, and OFM distance to compare the candidate structures' similarity with the target ground truth structure. 
For Pb$_4$O$_4$ and V$_4$O$_4$F$_4$, AlphaCrystal-I could not generate any structure candidates as the predicted space group was different from the ground truth, and the suggested template for these formulas contained many special coordinates (such as 0, 0.25, 0.33, 0.5, 0.66) that AlphaCrystal-I could not further relax.
For all other six formulas, AlphaCrystal-II was able to provide the structures that had the lower energy distance, fingerprint distance, and OFM distance with the ground truth structure. When considering Sinkhorn distance, AlphaCrystal-I only outperformed AlphaCrystal-II once on Fe$_4$As$_4$Se$_4$. For V$_2$Cl$_{10}$ and Mn$_4$Cu$_4$P$_4$, the structure candidates generated by AlphaCrystal-I had lower superpose distance.
In general, AlphaCrystal-II outperformed AlphaCrystal-I, especially when the predicted structure contained special coordinates.

\begin{table}[h]
\begin{center}
\caption{Performance Comparison of AlphaCrystal-I and AlphaCrystal-II.}
\label{tab:compareI_and_II}
\begin{tabular}{|l|l|r|r|r|r|r|}
\hline
Formula                   & Model & \multicolumn{1}{l|}{\begin{tabular}[c]{@{}l@{}}energy \\ distance (eV)\end{tabular}} & \multicolumn{1}{l|}{\begin{tabular}[c]{@{}l@{}}sinkhorn \\ distance(Å)\end{tabular}} & \multicolumn{1}{l|}{\begin{tabular}[c]{@{}l@{}}superpose \\ distance(Å)\end{tabular}} & \multicolumn{1}{l|}{\begin{tabular}[c]{@{}l@{}}fingerprint \\ distance\end{tabular}} & \multicolumn{1}{l|}{\begin{tabular}[c]{@{}l@{}}OFM \\ distance\end{tabular}} \\ \hline
\multirow{2}{*}{Pb$_4$O$_4$}     & AlphaCrystal-I   & None &None    & None & None     & None     \\ \cline{2-7} 
& AlphaCrystal-II  & \textbf{0.0481}  & \textbf{13.8642}   & \textbf{0.7667}  & \textbf{0.3945}  & \textbf{0.0355}\\ \hline
\multirow{2}{*}{Co$_4$P$_8$}  & AlphaCrystal-I   & 5.5616   & 20.2276 & 1.0251 
& 1.2315    & 0.1106   \\ \cline{2-7} 
& AlphaCrystal-II  & \textbf{0.0011} & \textbf{17.3153}  & \textbf{0.9784}  & \textbf{1.0926}  & \textbf{0.0844}  \\ \hline
\multirow{2}{*}{Ir$_4$N$_8$} & AlphaCrystal-I   & 18.5385  & 17.2003  & 1.0512 
& 1.1572   & 0.2728  \\ \cline{2-7} 
& AlphaCrystal-II  & \textbf{0.4584}   & \textbf{14.2731}  & \textbf{0.9165} & \textbf{0.6723} & \textbf{0.0719} \\ \hline
\multirow{2}{*}{V$_2$Cl{$_1$$_0$}}  & AlphaCrystal-I   & 7.34  & 39.5468 & \textbf{1.5859} & 1.3944  & 0.0817  \\ \cline{2-7} 
& AlphaCrystal-II  & \textbf{0.1806}  & \textbf{33.2543} & 1.6411& \textbf{1.1116} & \textbf{0.0357} \\ \hline
\multirow{2}{*}{Co$_2$As$_2$S$_2$}  & AlphaCrystal-I   & 3.0549  & 9.5731 & 0.9939        & 1.9771   & 0.2249 \\ \cline{2-7} 
& AlphaCrystal-II  & \textbf{0.0326} & \textbf{7.7433}  & \textbf{0.9477}   & \textbf{0.1892} & \textbf{0.0279}  \\ \hline
\multirow{2}{*}{V$_4$O$_4$F$_4$}    & AlphaCrystal-I   & None &None    & None & None     & None     \\ \cline{2-7} 
& AlphaCrystal-II  & \textbf{0.0495}  & \textbf{15.2005}   & \textbf{1.033}  & \textbf{0.2289}  & \textbf{0.0355}  \\ \hline
\multirow{2}{*}{Fe$_4$As$_4$Se$_4$} & AlphaCrystal-I   & 3.4682 & \textbf{16.0662} & 1.3837 
& 1.262  & 0.1906 \\ \cline{2-7} 
& AlphaCrystal-II  & \textbf{0.006} & 17.341 & \textbf{1.0916} & \textbf{0.8575} & \textbf{0.0655} \\ \hline
\multirow{2}{*}{Mn$_4$Cu$_4$P$_4$}  & AlphaCrystal-I   & 11.0241  & 20.0757  & \textbf{1.0341} & 1.2256  & 0.2327  \\ \cline{2-7} 
& AlphaCrystal-II  & \textbf{0.0168} & \textbf{19.4467} & 1.1193 & \textbf{0.8246}  & \textbf{0.1476} \\ \hline
\multirow{2}{*}{\# of best}  & AlphaCrystal-I   & 0  & \textbf{1}  & \textbf{2} &  0 & 0\\ \cline{2-7} 
& AlphaCrystal-II  & \textbf{8} & \textbf{7}  & \textbf{6}  & \textbf{8} & \textbf{8}\\ \hline
\end{tabular}
\end{center}
\end{table}

\subsubsection{Performance of distance matrix prediction in AlphaCrystal-II}

To assess the strengths and weakness of our AlphaCrystal-II model, we evaluated whether the deep neural network of our AlphaCrystal-II model can predict the distance matrix with good performance by learning the atomic pair relationships from the chemical compositions. We evaluated the performance with different experiments with varying hyper-parameters, including the type of samples, the number of discrete groups, and the sizes of training and test datasets. The distance matrix prediction performance is summarized in Table \ref{tab:performances}. In the training of these distance matrix prediction models, we set the number of epochs to 125 and utilized the Adam optimizer with a learning rate of 0.001.

Table \ref{tab:performances} includes distance matrix prediction performance metrics including MSE, RMSE, and distance map overlap. It also summarizes the results of multiple comparison trials. Firstly, comparing different sizes of training datasets, there is a clear trend of decreasing MSE and RMSE with more training samples. For example, for the AlphaCrystal-II model with a discretization group number of 5, when the number of training samples is increased from 1,000 to 10,000, the RMSE is reduced by 53.57\%. Secondly, we set different numbers of discrete groups for the basic AlphaCrystal-II model. Comparing groups 5, 10, and 20 with 10,000 training samples, RMSE is reduced by 11.69\% and 18.88\%, respectively, indicating better performance when a larger distance encoding group is used. 
In order to explore whether focusing on the same type of materials has a better performance in distance matrix prediction, we trained three special models for distance matrix prediction for three families of materials: AlphaCrystal-II\_cubic, AlphaCrystal-II\_binary, and AlphaCrystal-II\_ternary. AlphaCrystal-II\_cubic model is trained by materials whose crystal system is cubic. AlphaCrystal-II\_binary and AlphaCrystal-II\_ternary are trained by binary and ternary materials. As shown in Table \ref{tab:performances} (last three rows), the AlphaCrystal-II\_cubic achieved the smallest MSE (1.0125 Å) and RMSE (0.7394 Å) compared to all other models, indicating that training a specialized distance matrix prediction model is beneficial to improve their performance. For example, one may want to train such a model for perovskite materials. Next, we found the AlphaCrystal-II\_ternary model achieved the second-best performance out of these three specialized models with an MSE of 2.0832 Å and RMSE of 1.2779 Å, together with the highest distance map overlap. This can be understood by the fact that the ternary materials have the maximum number of training samples and many of them share some limited number of structure prototypes, which makes it easier to predict their distance matrices. Finally, the AlphaCrystal-II\_binary model achieved the worst performance despite its reasonable number of training samples (4,678) with an MSE of 2.7835  Å and RMSE of 1.3109, which can be partially attributed to the fact that binary structures are more diverse in terms of structural prototypes. Overall, looking at the distance map overlap performance criterion, we found all three models have achieved reasonably good results with scores ranging from 0.9746 to 0.9985.

\begin{table}[h]
\begin{center}
\caption{Distance matrix prediction performance of AlphaCrystal-II}
\label{tab:performances}
\begin{tabular}{|c|>{\centering\arraybackslash}p{2cm}|c|c|c|c|c|}
\hline
Model        &  Distance encoding groups  & Training dataset  & Test dataset & MSE & RMSE & Overlap  \\ \hline
AlphaCrystal-II  & 5    & 1,000   &  100   & 4.9538 & 1.8841 & 0.9045\\ \hline
AlphaCrystal-II & 5      &10,000   &  1,000    & 2.3764 & 1.2269 & 0.9979\\ \hline
AlphaCrystal-II & 10      &1,000    &   100  & 2.8586 & 1.3399 & 0.9981\\ \hline
AlphaCrystal-II & 10      &10,000   &   1,000  & 1.8932 & 1.0984 & 0.9989 \\ \hline
AlphaCrystal-II & 20     &10,000   &   1,000  &  1.5591 & 0.9239 & 0.9992 \\ \hline
AlphaCrystal-II\_cubic & 10 & 2,638 & 200  & 1.0125 & 0.7394  &  0.9746\\ \hline
AlphaCrystal-II\_binary & 10  & 4,678 & 400  & 2.7835 & 1.3109 &0.9980  \\ \hline
AlphaCrystal-II\_ternary & 10  & 9,292 & 1,000 & 2.0832 & 1.2779  & 0.9985  \\ \hline
\end{tabular}
\end{center}
\end{table}

\begin{figure}[htb!]
	\centering
	\begin{subfigure}{.46\textwidth}
		\includegraphics[width=\textwidth]{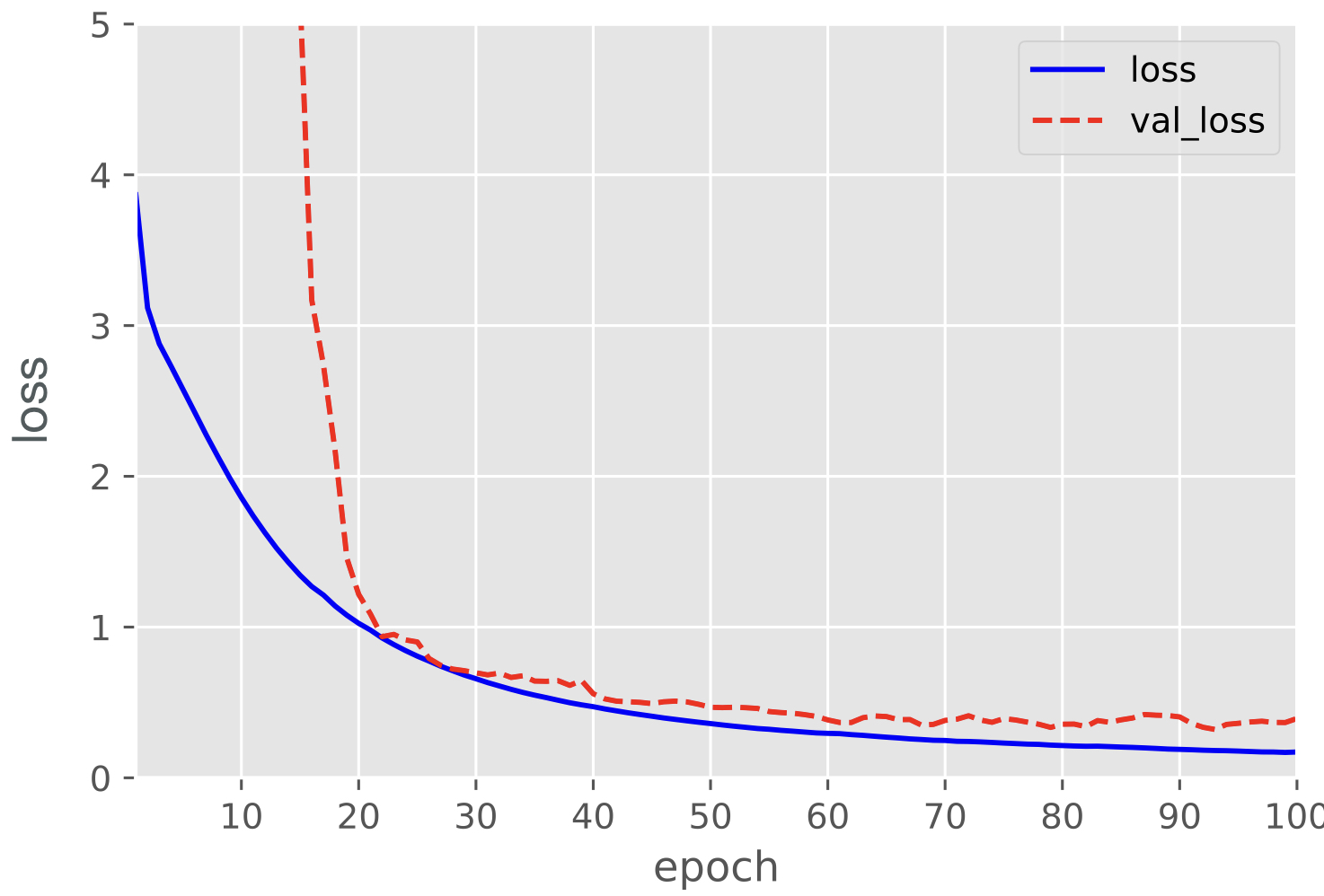}
        \caption{1,000 training samples}
        \label{fig:loss_1000}
	\end{subfigure}
	\begin{subfigure}{.48\textwidth}
		\includegraphics[width=\textwidth]{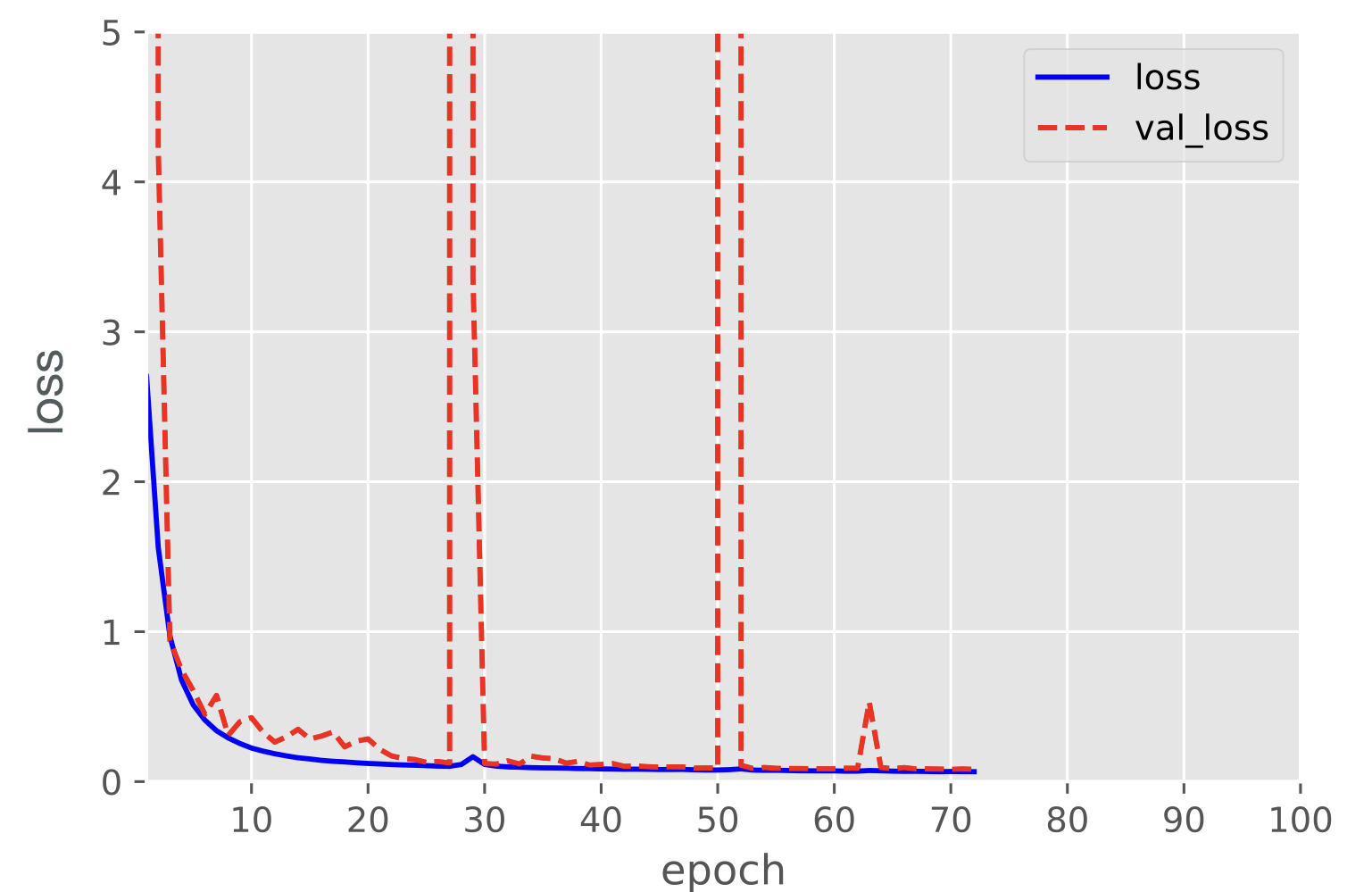}
		\caption{10,000 training samples}
		\label{fig:loss_10000}
 \vspace{3pt}
	\end{subfigure}
	\caption{Training and validation loss curves of the residual network model for distance matrix prediction. (a) the training curve of the AlphaCrystal-II model with 1,000 training samples; (b) the training curve of the AlphaCrystal-II model with 10,000 training samples.}
	\label{fig:loss}
\end{figure}

To further understand the prediction capability of our distance matrix predictor, we showed the training loss and validation loss of the neural network models from the AlphaCrystal-II models with 1,000 and 10,000 training samples in Figure \ref{fig:loss}. The discretization group number of both is set as 10. As the number of epochs increases, both the training loss and validation loss gradually decrease, when the loss curves hit a plateau, around 80 and 60 epochs, respectively, and they do not decrease any further and become stabilized. In Figure \ref{fig:loss_1000}, the loss curve shows a slow but steady decrease over time, indicating that the model is learning at a consistent rate. In Figure \ref{fig:loss_10000}, the loss curves show a steep drop during the first few training epochs, indicating that the model is learning quickly. The shapes of both training and validation loss curves demonstrated that our distance matrix network models were well-trained.

\begin{figure}[htb!]
	\centering

 	\begin{subfigure}{0.98\textwidth}
		\includegraphics[width=\textwidth]{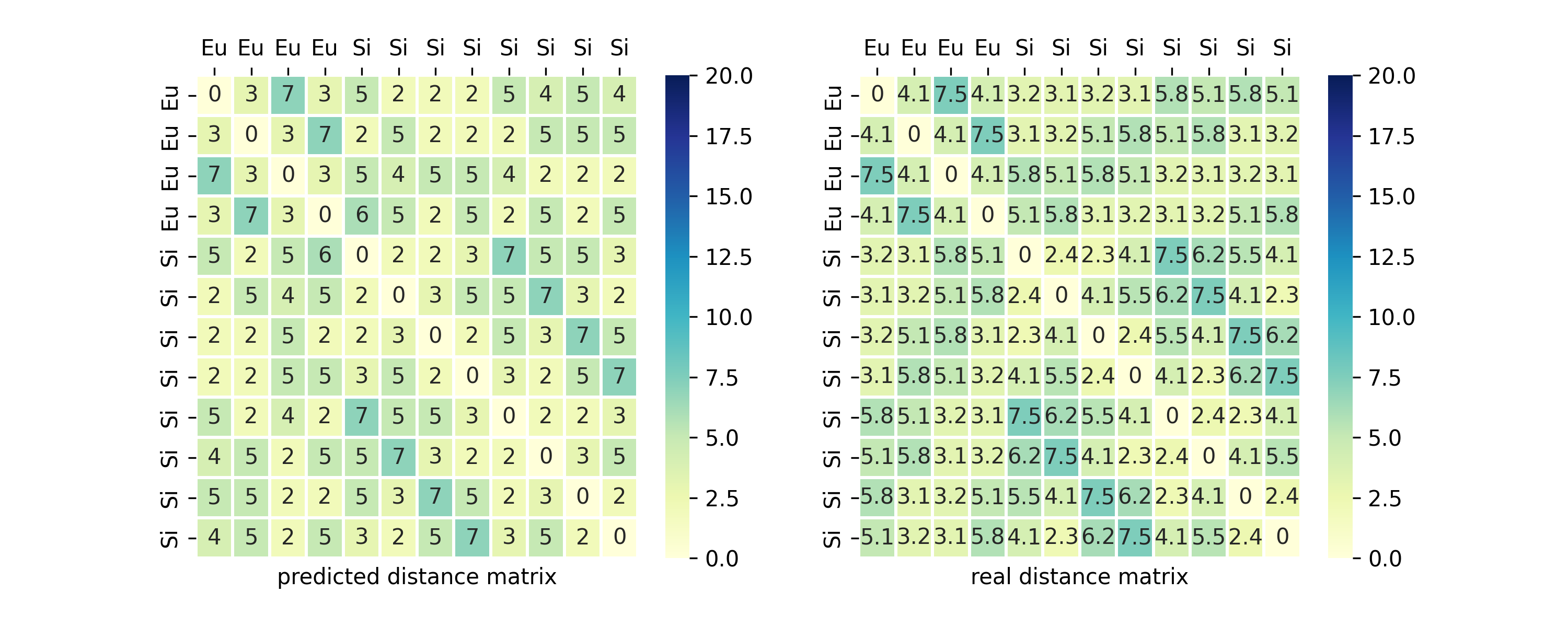}
        \caption{EuSi$_{2}$}
	\end{subfigure}
	
	\begin{subfigure}{0.98\textwidth}
		\includegraphics[width=\textwidth]{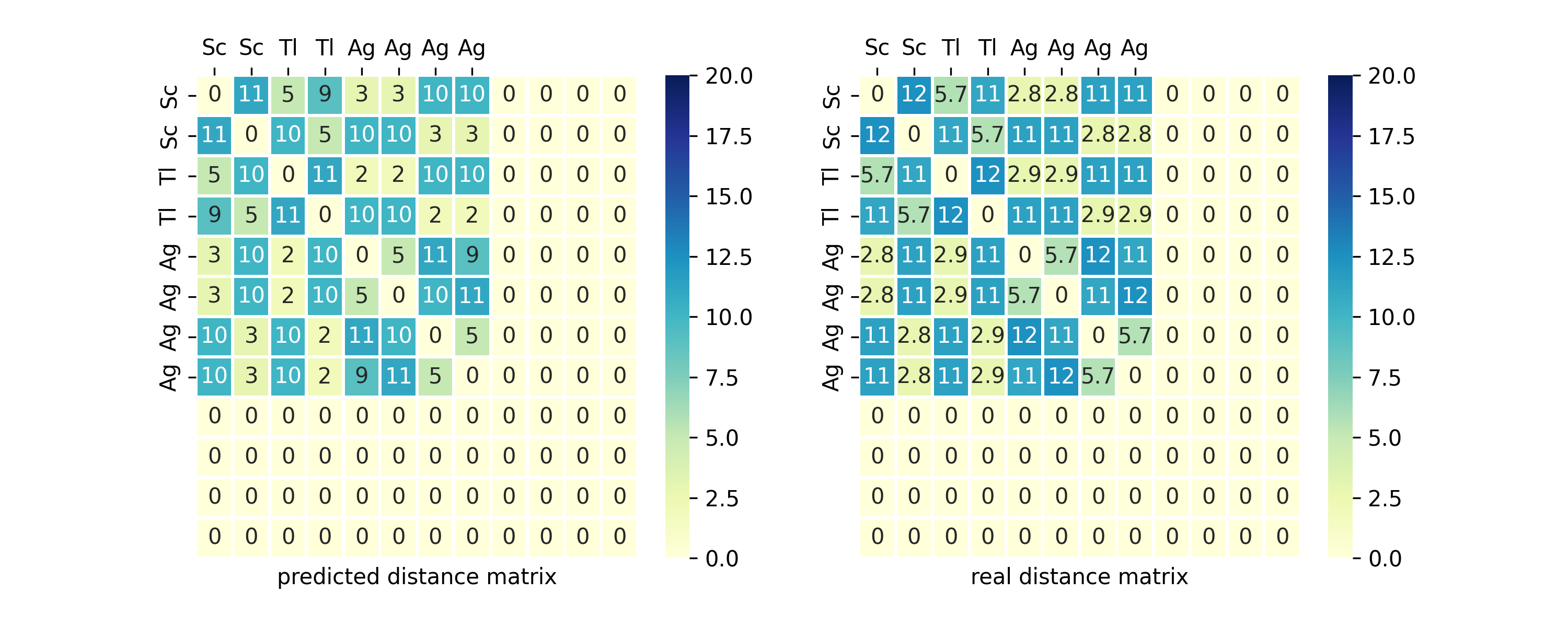}
        \caption{ScTlAg$_{2}$}
	\end{subfigure}
	\caption{Comparison of the predicted and real distance matrix with the discretization group number of 20. (a) EuSi$_{2}$(mpid:1072248), (b) ScTlAg$_{2}$ (mpid:1093619).}
	\label{fig:predicted_and_real_materix_10}
\end{figure}

To obtain an intuitive understanding of the distance matrix, Figure \ref{fig:predicted_and_real_materix_10} shows two examples of the real ground truth distance matrix and the predicted distance matrix for two compounds EuSi$_{2}$ and ScTlAg$_{2}$ for which the distance discretization group number is set to 20. The highlighted cells show the accurately predicted inter-atomic interaction distances. It can be found that both cases have relatively high accuracy in terms of distance matrix cell matching. These figures demonstrate that our deep neural network models can learn the pair-wise relationship between atoms for distance matrix prediction. We also find that as the number of discretized distance groups increases within a reasonable range, the predicted distance labels are more accurate.

\subsubsection{Performance of distance matrix based crystal structure reconstruction}

Another major module of our AlphaCrystal-II algorithm is the distance matrix based crystal structure reconstruction. For each query composition, we first predict their atomic distance matrix from their composition and then use the DMCrystal algorithm \cite{hu2020distance} to predict its three-dimensional crystal structure. For each query formula, we generate 50 candidate structures using the genetic algorithm based DMCrystal algorithm for structure reconstruction. We then use the Bayesian optimization routine of the M3GNet package \cite{chen2022universal} to relax these structures and predict their formation energies. 
We summarize the predicted RMSE performances of 10 binary materials, 10 ternary materials, and 10 materials with more than 4 elements in Table \ref{tab:perf_compare}.  "Mp\_id" represents the specific ID number of each chemical formula in the material project database. "Distance Matrix RMSE" indicates the RMSE values of our predicted distance matrix. The row of "AlphaCrystal-II Top-10 RMSE" lists the lowest RMSE values within the 10 lowest formation energies while the row of "AlphaCrystal-II Top-20 RMSE" lists the lowest RMSE values within the 20 lowest formation energies.

\section{Conclusion}
\label{sec:others}
We present AlphaCrystal-II, a novel data-driven deep learning approach for crystal structure prediction that exploits the abundant inter-atomic interaction patterns present in known crystal material structures. Our method first employs a deep residual network to learn and predict the atomic distance matrix for a given material composition. It then utilizes this matrix to reconstruct the 3D crystal structures through a genetic algorithm. Subsequently, we employ the Bayesian Optimization algorithm of the M3GNET package to relax these structures and calculate their formation energies, identifying the most stable configurations.

Through extensive experiments, we demonstrate that our model effectively captures the implicit inter-atomic relationships and showcases its effectiveness and reliability in leveraging such information for crystal structure prediction. Our results indicate that AlphaCrystal-II can successfully reconstruct the crystal structures of a wide range of materials, including binary, ternary, and multi-component systems with more than four elements. By harnessing the abundant inter-atomic interaction patterns, our data-driven, knowledge-guided approach paves the way for large-scale crystal structure prediction in the near future.

This work highlights the potential of deep learning and data-driven methods in accelerating the discovery and design of new materials with tailored properties. By combining data-driven approaches with domain knowledge and physics-based methods, we can unlock new avenues for materials research and development, ultimately contributing to the advancement of various technological domains.

\section{Availability of data}

The data that support the findings of this study are openly available in the Materials Project database at https://legacy.materialsproject.org/ 

\section{Contribution}
Conceptualization, J.H.; methodology, J.H. and Y.S.; software, Y.S., and J.H; validation, Y.S. E.S. J.H.;  investigation, Y.S., L.W., Q.L., J.H., ; resources, J.H.; data curation, J.H., Y.S.; writing--original draft preparation, Y.S., R.D., L.W., J.H.; writing--review and editing, J.H, R.D.; visualization, Y.S. J.H.; supervision, J.H.;  funding acquisition, J.H.

\section{Acknowledgement}
Research reported in this work was supported in part by NSF under grants 1940099 and 1905775 and by NSF SC EPSCoR Program under award number (NSF Award OIA-1655740 and GEAR-CRP 19-GC02). The views, perspectives, and content do not necessarily represent the official views of the SC EPSCoR Program or those of the NSF. 

\bibliographystyle{unsrt}  
\bibliography{references}  

\begin{thebibliography}{10}

\bibitem{oganov2019structure}
Artem~R Oganov, Chris~J Pickard, Qiang Zhu, and Richard~J Needs.
\newblock Structure prediction drives materials discovery.
\newblock {\em Nature Reviews Materials}, 4(5):331--348, 2019.

\bibitem{zunger2018inverse}
Alex Zunger.
\newblock Inverse design in search of materials with target functionalities.
\newblock {\em Nature Reviews Chemistry}, 2(4):1--16, 2018.

\bibitem{kim2020inverse}
Baekjun Kim, Sangwon Lee, and Jihan Kim.
\newblock Inverse design of porous materials using artificial neural networks.
\newblock {\em Science advances}, 6(1):eaax9324, 2020.

\bibitem{dan2019generative}
Yabo Dan, Yong Zhao, Xiang Li, Shaobo Li, Ming Hu, and Jianjun Hu.
\newblock Generative adversarial networks (gan) based efficient sampling of
  chemical space for inverse design of inorganic materials.
\newblock {\em arXiv preprint arXiv:1911.05020}, 2019.

\bibitem{bradshaw2019model}
John Bradshaw, Brooks Paige, Matt~J Kusner, Marwin Segler, and Jos{\'e}~Miguel
  Hern{\'a}ndez-Lobato.
\newblock A model to search for synthesizable molecules.
\newblock In {\em Advances in Neural Information Processing Systems}, pages
  7937--7949, 2019.

\bibitem{noh2019inverse}
Juhwan Noh, Jaehoon Kim, Helge~S Stein, Benjamin Sanchez-Lengeling, John~M
  Gregoire, Alan Aspuru-Guzik, and Yousung Jung.
\newblock Inverse design of solid-state materials via a continuous
  representation.
\newblock {\em Matter}, 1(5):1370--1384, 2019.

\bibitem{ren2020inverse}
Zekun Ren, Juhwan Noh, Siyu Tian, Felipe Oviedo, Guangzong Xing, Qiaohao Liang,
  Armin Aberle, Yi~Liu, Qianxiao Li, Senthilnath Jayavelu, et~al.
\newblock Inverse design of crystals using generalized invertible
  crystallographic representation.
\newblock {\em arXiv preprint arXiv:2005.07609}, 2020.

\bibitem{glass2006uspex}
Colin~W Glass, Artem~R Oganov, and Nikolaus Hansen.
\newblock Uspex—evolutionary crystal structure prediction.
\newblock {\em Computer physics communications}, 175(11-12):713--720, 2006.

\bibitem{oganov2011modern}
Artem~R Oganov.
\newblock {\em Modern methods of crystal structure prediction}.
\newblock John Wiley \& Sons, 2011.

\bibitem{kvashnin2019computational}
Alexander~G Kvashnin, Zahed Allahyari, and Artem~R Oganov.
\newblock Computational discovery of hard and superhard materials.
\newblock {\em Journal of Applied Physics}, 126(4):040901, 2019.

\bibitem{maddox1988crystals}
John Maddox.
\newblock Crystals from first principles.
\newblock {\em Nature}, 335(6187):201--201, 1988.

\bibitem{woodley2008crystal}
Scott~M Woodley and Richard Catlow.
\newblock Crystal structure prediction from first principles.
\newblock {\em Nature materials}, 7(12):937--946, 2008.

\bibitem{lyakhov2013new}
Andriy~O Lyakhov, Artem~R Oganov, Harold~T Stokes, and Qiang Zhu.
\newblock New developments in evolutionary structure prediction algorithm
  uspex.
\newblock {\em Computer Physics Communications}, 184(4):1172--1182, 2013.

\bibitem{xie2018crystal}
Tian Xie and Jeffrey~C Grossman.
\newblock Crystal graph convolutional neural networks for an accurate and
  interpretable prediction of material properties.
\newblock {\em Physical review letters}, 120(14):145301, 2018.

\bibitem{omee2022scalable}
Sadman~Sadeed Omee, Steph-Yves Louis, Nihang Fu, Lai Wei, Sourin Dey, Rongzhi
  Dong, Qinyang Li, and Jianjun Hu.
\newblock Scalable deeper graph neural networks for high-performance materials
  property prediction.
\newblock {\em Patterns}, 3(5), 2022.

\bibitem{wei2022crystal}
Lai Wei, Qinyang Li, Yuqi Song, Stanislav Stefanov, Edirisuriya Siriwardane,
  Fanglin Chen, and Jianjun Hu.
\newblock Crystal transformer: Self-learning neural language model for
  generative and tinkering design of materials.
\newblock {\em arXiv preprint arXiv:2204.11953}, 2022.

\bibitem{oganov2006crystal}
Artem~R Oganov and Colin~W Glass.
\newblock Crystal structure prediction using ab initio evolutionary techniques:
  Principles and applications.
\newblock {\em The Journal of chemical physics}, 124(24):244704, 2006.

\bibitem{wang2015materials}
Yanchao Wang, Jian Lv, Li~Zhu, Shaohua Lu, Ketao Yin, Quan Li, Hui Wang, Lijun
  Zhang, and Yanming Ma.
\newblock Materials discovery via calypso methodology.
\newblock {\em Journal of Physics: Condensed Matter}, 27(20):203203, 2015.

\bibitem{oganov2011evolutionary}
Artem~R Oganov, Andriy~O Lyakhov, and Mario Valle.
\newblock How evolutionary crystal structure prediction works and why.
\newblock {\em Accounts of chemical research}, 44(3):227--237, 2011.

\bibitem{wang2020calypso}
Yanchao Wang, Jian Lv, Quan Li, Hui Wang, and Yanming Ma.
\newblock Calypso method for structure prediction and its applications to
  materials discovery.
\newblock {\em Handbook of Materials Modeling: Applications: Current and
  Emerging Materials}, pages 2729--2756, 2020.

\bibitem{zhang2017materials}
Lijun Zhang, Yanchao Wang, Jian Lv, and Yanming Ma.
\newblock Materials discovery at high pressures.
\newblock {\em Nature Reviews Materials}, 2(4):1--16, 2017.

\bibitem{pretti2020symmetry}
Evan Pretti, Vincent~K Shen, Jeetain Mittal, and Nathan~A Mahynski.
\newblock Symmetry-based crystal structure enumeration in two dimensions.
\newblock {\em The Journal of Physical Chemistry A}, 124(16):3276--3285, 2020.

\bibitem{podryabinkin2019accelerating}
Evgeny~V Podryabinkin, Evgeny~V Tikhonov, Alexander~V Shapeev, and Artem~R
  Oganov.
\newblock Accelerating crystal structure prediction by machine-learning
  interatomic potentials with active learning.
\newblock {\em Physical Review B}, 99(6):064114, 2019.

\bibitem{agrawal2019deep}
Ankit Agrawal and Alok Choudhary.
\newblock Deep materials informatics: Applications of deep learning in
  materials science.
\newblock {\em MRS Communications}, 9(3):779--792, 2019.

\bibitem{louis2020graph}
Steph-Yves Louis, Yong Zhao, Alireza Nasiri, Xiran Wang, Yuqi Song, Fei Liu,
  and Jianjun Hu.
\newblock Graph convolutional neural networks with global attention for
  improved materials property prediction.
\newblock {\em Physical Chemistry Chemical Physics}, 22(32):18141--18148, 2020.

\bibitem{song2021computational}
Yuqi Song, Edirisuriya M~Dilanga Siriwardane, Yong Zhao, and Jianjun Hu.
\newblock Computational discovery of new 2d materials using deep learning
  generative models.
\newblock {\em ACS Applied Materials \& Interfaces}, 13(45):53303--53313, 2021.

\bibitem{kim2021deep}
Yongtae Kim, Youngsoo Kim, Charles Yang, Kundo Park, Grace~X Gu, and Seunghwa
  Ryu.
\newblock Deep learning framework for material design space exploration using
  active transfer learning and data augmentation.
\newblock {\em npj Computational Materials}, 7(1):1--7, 2021.

\bibitem{ryan2018crystal}
Kevin Ryan, Jeff Lengyel, and Michael Shatruk.
\newblock Crystal structure prediction via deep learning.
\newblock {\em Journal of the American Chemical Society}, 140(32):10158--10168,
  2018.

\bibitem{cheng2022crystal}
Guanjian Cheng, Xin-Gao Gong, and Wan-Jian Yin.
\newblock Crystal structure prediction by combining graph network and
  optimization lgorithm.
\newblock {\em Nature communications}, 13(1):1--8, 2022.

\bibitem{CSPML}
Minoru Kusaba, Chang Liu, and Ryo Yoshida.
\newblock Crystal structure prediction with machine learning-based element
  substitution.
\newblock {\em Computational Materials Science}, 211:111496, 2022.

\bibitem{wei2022tcsp}
Lai Wei, Nihang Fu, Edirisuriya~MD Siriwardane, Wenhui Yang, Sadman~Sadeed
  Omee, Rongzhi Dong, Rui Xin, and Jianjun Hu.
\newblock Tcsp: a template-based crystal structure prediction algorithm for
  materials discovery.
\newblock {\em Inorganic Chemistry}, 2022.

\bibitem{jumper2021highly}
John Jumper, Richard Evans, Alexander Pritzel, Tim Green, Michael Figurnov,
  Olaf Ronneberger, Kathryn Tunyasuvunakool, Russ Bates, Augustin
  {\v{Z}}{\'\i}dek, Anna Potapenko, et~al.
\newblock Highly accurate protein structure prediction with alphafold.
\newblock {\em Nature}, 596(7873):583--589, 2021.

\bibitem{wei2019protein}
Guo-Wei Wei.
\newblock Protein structure prediction beyond alphafold.
\newblock {\em Nature Machine Intelligence}, 1(8):336--337, 2019.

\bibitem{adhikari2018dncon2}
Badri Adhikari, Jie Hou, and Jianlin Cheng.
\newblock Dncon2: improved protein contact prediction using two-level deep
  convolutional neural networks.
\newblock {\em Bioinformatics}, 34(9):1466--1472, 2018.

\bibitem{emerson2017protein}
Isaac~Arnold Emerson and Arumugam Amala.
\newblock Protein contact maps: A binary depiction of protein 3d structures.
\newblock {\em Physica A: Statistical Mechanics and its Applications},
  465:782--791, 2017.

\bibitem{kuhlman2019advances}
Brian Kuhlman and Philip Bradley.
\newblock Advances in protein structure prediction and design.
\newblock {\em Nature Reviews Molecular Cell Biology}, 20(11):681--697, 2019.

\bibitem{ingraham2019generative}
John Ingraham, Vikas Garg, Regina Barzilay, and Tommi Jaakkola.
\newblock Generative models for graph-based protein design.
\newblock In {\em Advances in Neural Information Processing Systems}, pages
  15820--15831, 2019.

\bibitem{senior2020improved}
Andrew~W Senior, Richard Evans, John Jumper, James Kirkpatrick, Laurent Sifre,
  Tim Green, Chongli Qin, Augustin {\v{Z}}{\'\i}dek, Alexander~WR Nelson, Alex
  Bridgland, et~al.
\newblock Improved protein structure prediction using potentials from deep
  learning.
\newblock {\em Nature}, 577(7792):706--710, 2020.

\bibitem{hu2023deep}
Jianjun Hu, Yong Zhao, Qin Li, Yuqi Song, Rongzhi Dong, Wenhui Yang, and
  Edirisuriya~MD Siriwardane.
\newblock Deep learning-based prediction of contact maps and crystal structures
  of inorganic materials.
\newblock {\em ACS omega}, 8(29):26170--26179, 2023.

\bibitem{jain2013commentary}
Anubhav Jain, Shyue~Ping Ong, Geoffroy Hautier, Wei Chen, William~Davidson
  Richards, Stephen Dacek, Shreyas Cholia, Dan Gunter, David Skinner, Gerbrand
  Ceder, et~al.
\newblock Commentary: The materials project: A materials genome approach to
  accelerating materials innovation.
\newblock {\em APL materials}, 1(1), 2013.

\bibitem{li2021mlatticeabc}
Yuxin Li, Wenhui Yang, Rongzhi Dong, and Jianjun Hu.
\newblock Mlatticeabc: generic lattice constant prediction of crystal materials
  using machine learning.
\newblock {\em ACS omega}, 6(17):11585--11594, 2021.

\bibitem{hu2020distance}
Jianjun Hu, Wenhui Yang, and Edirisuriya~M Dilanga~Siriwardane.
\newblock Distance matrix-based crystal structure prediction using evolutionary
  algorithms.
\newblock {\em The Journal of Physical Chemistry A}, 124(51):10909--10919,
  2020.

\bibitem{chen2022universal}
Chi Chen and Shyue~Ping Ong.
\newblock A universal graph deep learning interatomic potential for the
  periodic table.
\newblock {\em arXiv preprint arXiv:2202.02450}, 2022.

\bibitem{housecroft2012inorganic}
Catherine~E. Housecroft and Alan~G. Sharpe.
\newblock {\em Inorganic Chemistry}.
\newblock Pearson Education Limited, 4 edition, 2012.

\bibitem{villars2004data}
P~Villars, K~Cenzual, J~Daams, Y~Chen, and S~Iwata.
\newblock Data-driven atomic environment prediction for binaries using the
  mendeleev number: Part 1. composition ab.
\newblock {\em Journal of alloys and compounds}, 367(1-2):167--175, 2004.

\bibitem{schwerdtfeger20192018}
Peter Schwerdtfeger and Jeffrey~K Nagle.
\newblock 2018 table of static dipole polarizabilities of the neutral elements
  in the periodic table.
\newblock {\em Molecular Physics}, 117(9-12):1200--1225, 2019.

\bibitem{he2016identity}
Kaiming He, Xiangyu Zhang, Shaoqing Ren, and Jian Sun.
\newblock Identity mappings in deep residual networks.
\newblock In {\em European conference on computer vision}, pages 630--645.
  Springer, 2016.

\bibitem{wei2024towards}
Lai Wei, Qin Li, Sadman~Sadeed Omee, and Jianjun Hu.
\newblock Towards quantitative evaluation of crystal structure prediction
  performance.
\newblock {\em Computational Materials Science}, 235:112802, 2024.

\bibitem{ong2013python}
Shyue~Ping Ong, William~Davidson Richards, Anubhav Jain, Geoffroy Hautier,
  Michael Kocher, Shreyas Cholia, Dan Gunter, Vincent~L Chevrier, Kristin~A
  Persson, and Gerbrand Ceder.
\newblock Python materials genomics (pymatgen): A robust, open-source python
  library for materials analysis.
\newblock {\em Computational Materials Science}, 68:314--319, 2013.

\bibitem{chen2019graph}
Chi Chen, Weike Ye, Yunxing Zuo, Chen Zheng, and Shyue~Ping Ong.
\newblock Graph networks as a universal machine learning framework for
  molecules and crystals.
\newblock {\em Chemistry of Materials}, 31(9):3564--3572, 2019.

\end{thebibliography}

\end{document}